\documentclass[a4,12pt]{article}
\usepackage{amsmath}
\usepackage{amssymb}
\usepackage{graphicx}
\usepackage{amsfonts}
\usepackage[sort&compress,square,numbers,comma]{natbib}
\usepackage{mathtools}
\usepackage{authblk}
\usepackage{enumitem}

\title{Hawking radiation of rotating BTZ Black hole based on modified dispersion relation and Rarita-Schwinger equation}

\author[1]{Sapam Gayatri Devi\footnote{Corresponding Author\\ sapamgayatri@manipuruniv.ac.in}}
\author[1]{I. Ablu Meitei}
\author[2]{T. Ibungochouba Singh}
\author[3]{Aheibam Keshwarjit Singh}
\author[1]{K. Yugindro Singh}
\affil[1]{\small{{Department of Physics, Manipur University, Imphal-795003, India} }}
\affil[2]{Department of Mathematics, Manipur University, Imphal-795003, India}
\affil[3]{Department of Physics, DM college of Science , Imphal 795001, India}

\date{}

\begin{document}
\maketitle

\begin{abstract}
In this paper, tunneling of fermions from rotating BTZ black hole is investigated using modified dispersion relation (MDR) and Rarita-Schwinger equation. The effect of modified dispersion relation (MDR) on the tunneling of fermions rises the Hawking temperature of rotating BTZ black hole. It is observed that the modified Hawking temperature of the black hole depends not only on the radial parameters of the black hole but also on the angular parameters of the black hole and the coupling constant $\sigma$ . Further, the entropy and the heat capacity of the black hole are also studied. \\
\end{abstract}

\newpage

\section{Introduction}
In 1972, Bekenstein proposed that a black hole has an entropy that is proportional to the area of the event horizon of the black hole.$^{1,2}$ This leads to the foundation of Black hole thermodynamics. Soon after, in 1974, Hawking proposed that a black hole could emit particles which have thermal spectrum and these radiations are now, known as Hawking radiation.$^{3,4}$ Later, several approaches have been established to investigate the Hawking radiation of different black hole systems. Semi-classical tunneling approach provides an elementary concept to understand the Hawking radiation as a tunneling of particles across the event horizon of a black hole.$^{5,6}$ Hawking radiation is studied following the classical action of the particle by using Hamilton-Jacobi equation.\cite{7} This method is considered to be an extension of the complex path analysis.\cite{8} In both the methods, tunneling probability for the classically forbidden trajectory is calculated from inside to outside of event horizon of the black hole, which is given by,  
$\Gamma = \exp (\frac{-2}{\hbar}$Im $I)$, where $I$ is the action of the particle. 

According to quantum gravity theories, there exists a minimal observable length of the order of Planck length,$^{9-11}$ $l_{p}= \sqrt{\frac{G\hbar}{c^{3}}} \sim 10^{-33}cm$. Due to quantum gravity effects, the fundamental commutation relation gets modified and leads to Generalised Uncertainty Principle (GUP) which is given by, $\Delta x \Delta p \geq \frac{\hbar}{2} [1+ \beta (\Delta p)^{2}]$
where $\beta = \beta _{o} \frac{(l_{p})^{2}}{\hbar^{2}}$, $\ell_{p}$ is Planck length, $\beta_{o}$ is a dimensionless parameter of order unity.\cite{12} By using GUP in the study of Hawking radiation, many interesting results are obtained. Incorporating quantum gravity effects, Hawking radiation has been widely studied as a tunneling radiation of scalar particles and fermion particles across event horizon for various system of black holes.$^{13-26}$ Quantum gravity effects also slows down the increase in Hawking temperature and the corrected Hawking temperature depends not only on the properties of a black hole but also on the properties of the emitted particles such as mass, angular momentum and energy.$^{27-31}$ Theoretically, it has been proposed that quantum gravity effects can prevent total black hole evaporation and it can also study information loss paradox.$^{32-34}$

At the minimum observable length, the energy and momentum dispersion relation gets modified.$^{35-38}$ By applying modified Hamilton-Jacobi equation in higher dimensional charged AdS black hole in the investigation of tunneling of fermions, it is observed that the modified Hawking temperature of the black hole is higher than its original Hawking temperature. This leads to higher emission rate of particles following Stefan - Boltzmann law.\cite{39} This is due to the effect of Modified Dispersion Relation (MDR) that helps in boosting the emission of particles from the black hole. By using MDR in the presence of quantum gravity effects and semiclassical approximation method, fermion tunneling radiation for nonstationary symmetric black hole is studied and it is observed that the modified Hawking Radiation depends on both radial and angular parameter of the black hole.\cite{40} Similar result is also observed in the Demianski-Newman black hole at higher energy scales.\cite{41} The modified Hawking radiation for vector boson particle tunneling near the event horizon of stationary and non-stationary Kerr-Newmann de-Sitter black hole is also studied.\cite{42} It is observed that the modified Hawking radiation for both the stationary and non -stationary black hole are dependent on the black hole parameters such as mass and charge of the black hole, cosmological constant and on the retarded time. Modified tunneling radiation for fermions with arbitrary spins in Vaidya-Bonner black hole is studied.\cite{43} In this paper, we investigate quantum tunneling of fermions from rotating BTZ black hole by applying  MDR and Rarita-Schwinger equation. We also study some thermodynamic parameters of the black hole such as entropy and heat capacity. 

The outline of this paper is as follows. In section 2, the Rarita-Schwinger equation in curved spacetime is revisited. In section 3, the modified Hamilton-Jacobi equation is applied on the rotating BTZ black hole to investigate the modified Hawking temperature across the event horizon of the rotating BTZ black hole. In section 4, the modified Hawking temperature is used to find corrections to the entropy of the rotating BTZ black hole. In section 5, modified heat capacity of the rotating BTZ black hole is studied. Lastly, in section 6, we summarise our results with a brief discussion.   
\section{Review on modified Hamilton-Jacobi equation}
In quantum gravity theory at high energy scales, the relativistic dispersion relation gets modified at the order of Planck scale as$^{35-38}$
\begin{eqnarray}\label{1}
\xi^{2} = \overrightarrow{p}^{2} + m^{2} - (L\xi)^{\alpha} \overrightarrow{p}^{2},
\end{eqnarray}

where $\xi$ and $\overrightarrow{p}$ are the energy and momentum of the particle respectively, m is the mass of the particle and L is a constant of the magnitude of the Planck scale. Eq. (\ref{1}) is obtained from Liouville-string model when $\alpha=1 $. The modified Dirac equation of fermions is obtained by Kruglov in flat spacetime with $\alpha$ = 2 as\cite{44}
\begin{eqnarray}\label{2}
\left(\overline{\gamma}^{\mu} \partial_{\mu} + \frac{m}{\hbar} -i L \overline{\gamma}^{t} \partial_{t}  \overline{\gamma}^{a} \partial_{a}\right) \psi = 0.
\end{eqnarray}
where $\overline{\gamma}^{\mu}$ is the gamma matrix, $a$ and $\mu$ are space and spacetime coordinates respectively. The general form of equation of motion of fermions, corresponding to different spins represented by $\alpha_{k}$, in flat spacetime proposed by Rarita and Schwinger is\cite{45} 
\begin{eqnarray}\label{3}
\left(\overline { \gamma}^{\mu} \partial_{\mu} + \frac{m}{\hbar} \right) \psi_{\alpha 1\cdots \alpha_{k}} = 0.
\end{eqnarray}

Eq. (\ref{3}) is called Rarita-Schwinger equation and it obeys the condition
\begin{eqnarray}\label{4}
\overline{\gamma}^{\mu}  \psi_{\mu \alpha_{2}\cdots \alpha_{ k}}= \partial_{\mu} \psi^{\mu }_{\mu \alpha_{ 2}\cdots \alpha k}= \psi^{\mu}_{\mu \alpha_{3}\cdots \alpha_{k}}=0.
\end{eqnarray}
Using $\alpha=2$ in Eq. (\ref{1}), the Rarita-Schwinger equation in flat spacetime is\cite{44}
\begin{eqnarray}\label{5}
\left(\overline{\gamma}^{\mu} \partial_{\mu} + \frac{m}{\hbar} - \sigma \hbar \overline{\gamma}^{t} \partial_{t}  \overline{\gamma}^{a} \partial_{a}\right) \psi_{\alpha_{1}\cdots \alpha_{k}} = 0,
\end{eqnarray}
where $\sigma$ is the coupling constant such that $\sigma<< 1$ and the quantum scale correction term $\sigma \hbar \overline{\gamma}^{t} \partial_{t}  \overline{\gamma}^{a} \partial_{a}$ is infinitesimally small\cite{42}.
In curved spacetime, the gamma matrix become $\gamma^{\mu}$ and the partial derivative becomes covariant derivative $D_{\mu} $. Hence, the Rarita-Schwinger equation in curved spacetime is\cite{46}
\begin{eqnarray}\label{6}
\left(\gamma^{\mu} D_{\mu} + \frac{m}{\hbar} - \sigma \hbar \gamma^{t} D_{t} \gamma^{a} D_{a} \right) \psi_{\alpha_{1}\cdots \alpha_{k}} = 0,
\end{eqnarray}
where  $D_{\mu} \equiv \partial _{\mu} + \Omega_{\mu} + \frac{i}{\hbar} e A_{\mu}$, $\Omega_{\mu}$ is the spin connection in the curved spacetime, $\gamma^{\mu}$ satisfies the anti commutation relation $\lbrace\gamma^{\mu},\gamma^{\nu}\rbrace=2g^{\mu \nu}I$, and $eA_{\mu}$ is the charge term of Dirac equation.
Eq. (\ref{6}) satisfies the condition, 
\begin{eqnarray}\label{7}
\gamma^{\mu} \psi_{\mu \alpha_{ 2} \cdots \alpha_{k}} = D_{\mu} \psi^{\mu}_{\alpha_{2} \cdots \alpha_{ k}} = \psi^{\mu}_{\mu \alpha_{3} \cdots \alpha_{ k}} = 0.
\end{eqnarray}
In Eq. (\ref{6}), assuming coupling constant $\sigma \ll 1$ for small correction on quantum scale, $\sigma \hbar \overline{\gamma}^{t} \partial_{t}  \overline{\gamma}^{a} \partial_{a}$ becomes infinitesimal quantity. By considering the wave function for fermion tunneling radiation across the event horizon of a black hole as
\begin{eqnarray}\label{8}
\psi_{\alpha_{1}\cdots \alpha_{k}} = \xi_{\alpha_{1}\cdots \alpha_{k}} \exp \left( \frac{i}{\hbar} S\right),
\end{eqnarray}
where $\xi_{\alpha_{1}\cdots \alpha_{k}}$ is a matrix and S is the action of the fermion and it is defined as
\begin{eqnarray}\label{9}
S= -\omega t + S_{o}(\overrightarrow{r}, \theta , \phi),
\end{eqnarray}

with $\partial_{t} S$ = -$\omega$ and $\partial_{\phi} S= j$. Here, $\omega$ and $j$ are the radiant energy and the angular momentum of the emitted fermion particles. \\

Using Eq. (\ref{8}) in Eq. (\ref{6}) and neglecting higher order term of $\hbar$, we get
\begin{eqnarray}\label{10}
i \gamma^{\mu} (\partial_{\mu} S + eA_{\mu}) \xi_{\alpha{1}\cdots \alpha{k}}  + m \xi_{\alpha{1}\cdots \alpha{k}} - \sigma \gamma^{t}(\omega - eA_{t})\gamma^{a}(\partial_{a} S +eA_{a}) \xi_{\alpha{1}\cdots \alpha{k}}=0
\end{eqnarray}

Using the condition ($\partial_{\mu} S + eA_{\mu}$) = - $\gamma^{t}(\omega - eA_{t})+\gamma^{a}(\partial_{a} S +eA_{a})$ in Eq. (\ref{10}), the deformed Hamilton-Jacobi equation is obtained as 
\begin{eqnarray}\label{11}
g^{\mu\nu} (\partial_{\mu} S + eA_{\mu})  (\partial_{\nu} S + eA_{\nu}) + m^{2} - 2 \sigma m g^{tt} (\omega - e A_{t})^{2} =0
\end{eqnarray}

Eq. (\ref{11}) is called modified Hamilton-Jacobi equation or modified form of Rarita-Schwinger equation. The result of Eq. (\ref{11}) is not affected by specific spin and it can be used for any fermions in semiclassical approximation. When $e=0$, for uncharged rotating black hole, Eq. (\ref{11}) can be written as
\begin{eqnarray}\label{11a}
g^{\mu\nu} (\partial_{\mu} S) ( \partial_{\nu} S)  + m^{2} - 2 \sigma m g^{tt} \omega^{2} =0
\end{eqnarray} 
In the following section, we will discuss the fermion tunneling radiation near the event horizon of the BTZ black hole by using the modified Hamilton-Jacobi equation. 

\section{Tunneling Radiation of rotating BTZ black hole }
In this section, we investigate the tunneling radiation across the event horizon of rotating BTZ black hole based on MDR and by using modified Hamilton-Jacobi equation. The line element of rotating BTZ black hole is given as\cite{48,49}
\begin{eqnarray}\label{12}
ds^{2} = -f(r) dt^{2} + \frac{1}{g(r)} dr^{2} + r^{2} (d\phi - \frac{J}{2r^{2}} dt)^{2}, 
\end{eqnarray} 
where
\begin{eqnarray}\label{13}
 f(r) = g(r) = -M+ \frac{r^{2}}{\ell^{2}}+ \frac{J^{2}}{4r^{2}},
\end{eqnarray}
with M and J are ADM mass and angular momentum of rotating BTZ black hole respectively, $\ell$ is defined by $\Lambda$ = $\frac{-1}{\ell^{2}}$, $-\infty<t<\infty , 0<r<\infty $  and $ 0<\phi<2\pi $. From Eq. (\ref{12}), the two physically meaningful horizons of rotating BTZ black hole are\cite{24}
\begin{eqnarray}\label{14}
r_{\pm} = \ell \left[ \frac{M}{2} \left\lbrace 1\pm \sqrt{1- \left( \frac{J}{M\ell}\right) ^{2}}\right\rbrace \right] ^{1/2},
\end{eqnarray}
where $r_{+}$ is the outer event horizon and $r_{-}$ is the inner event horizon of rotating BTZ black hole. Here, the contravariant metric components of rotating BTZ black hole are
\begin{eqnarray}\label{15}
g^{11}&=&\frac{-1}{f(r)}, \cr
g^{13}&=&g^{31}=\frac{-J}{2r^{2}f(r)},\cr
g^{22}&=& g(r), \cr 
g^{33}&=& \frac{1}{r^{2}}-\frac{J^{2}}{4r^{4}f(r)}.
\end{eqnarray}

Using Eqs. (\ref{14}) and (\ref{15}) and  in Eq. (\ref{11a}), we obtained an equation of motion of fermions, 
\begin{eqnarray}\label{16}
\frac{-1}{f(r)}\omega^{2}+g(r)\partial_{r}S^{2}+(\frac{1}{r^{2}}-\frac{J^{2}}{4r^{4}f(r)})j^{2}+\frac{J}{2r^{2}f(r)}\omega j+m^{2}+2\sigma m\omega^{4}\frac{1}{f(r)} = 0.
\end{eqnarray}

Applying separation of variables in the Eq. (\ref{16}), we obtain the following solution of radial equation 
\begin{eqnarray}\label{17}
\partial _{r} S = \pm \frac{1}{g(r)}\left( \omega - j\Omega \right) \sqrt{1- \left\lbrace 2m\sigma\Phi^{2} 
+ \left( \frac{m^{2}g^{2}(r)-\frac{j^{2}g(r)}{r^{2}}}{(\omega - j \Omega)^{2}}\right) 
\right\rbrace },
\end{eqnarray}
where $\Omega = \frac{J}{2r^{2}}$ and $\Phi = \frac{\omega^{2}}{\omega-j\Omega}$. Now, The radial solution of the Eq. (\ref{17}) is 
\begin{eqnarray}\label{18}
S &=& \pm \int \frac{1}{g(r)}\left( \omega - j\Omega \right) \sqrt{1- \left\lbrace 2m\sigma\Phi^{2} 
+ \left( \frac{\frac{j^{2}g(r)}{r^{2}}+m^{2}g^{2}(r)}{(\omega - j \Omega)^{2}}\right) 
\right\rbrace } dr, \cr
&=& \pm \frac{ i \pi \ell^{2} r_{+}}{(r^{2}_{+}-r^{2}_{-})}  (\omega - j \Omega)(1-m\sigma \Phi^{2}),
\end{eqnarray}
where $\pm$ indicates for outgoing and incoming fermions. The tunneling rate across the event horizon of BTZ black hole is 
\begin{eqnarray}\label{19}
\Gamma = exp \left( \frac{-2\pi \ell^{2} r_{+} (\omega - j \Omega)}{r_{+}^{2} - r_{-}^{2}} (1-m\sigma \Phi^{2}) \right) = exp \left( \frac{-(\omega - j \Omega)}{T}\right), 
\end{eqnarray}

where T is the modified Hawking temperature and it is given by 
\begin{equation}\label{20}
T = \frac{r_{+}^{2} - r_{-}^{2}}{2\pi \ell^{2} r_{+} (1-m\sigma \Phi^{2})}  = T_{o} (1+m\sigma \Phi^{2}).
\end{equation}   
where $T_{o}$ is the original Hawking temperature of BTZ black hole. It can be seen clearly that the modified Hawking temperature increases due to the presence of the deformation parameter, $\sigma$. From Eq. (\ref{20}), it is observed that the Hawking temperature diverges when $\omega\longrightarrow j\Omega$ i.e, when the energy of the emitted particle is purely rotational. In the tunneling picture, when a pair of positive and negative energy particle is created just inside the event horizon, and the negative energy particle is attracted towards the center of the black hole, the positive energy particle gets radiated away radially due to conservation of linear momentum. So, the particle should possess non-rotational energy also. Hence, there should be a constraint $\omega \neq j\Omega$. Dependance on ($\omega-j \Omega$) and not on $\omega$ only may give rise to super-radiance.\cite{49,58} If $\sigma$ goes to zero, the original Hawking temperature is recovered. Also, the modified Hawking temperature depends on the energy and angular momentum of the emitted fermion particles. Figure 1 shows a schematic plots between Hawking temperature and mass of the black hole by assuming the the parameters in Eq. (\ref{20}) as $J=0.2$, $l=1$, $m=0.5$, $\sigma =0.2$ and $\Phi =2$. 

\begin{figure}[b]
\centerline{\includegraphics[width=15cm]{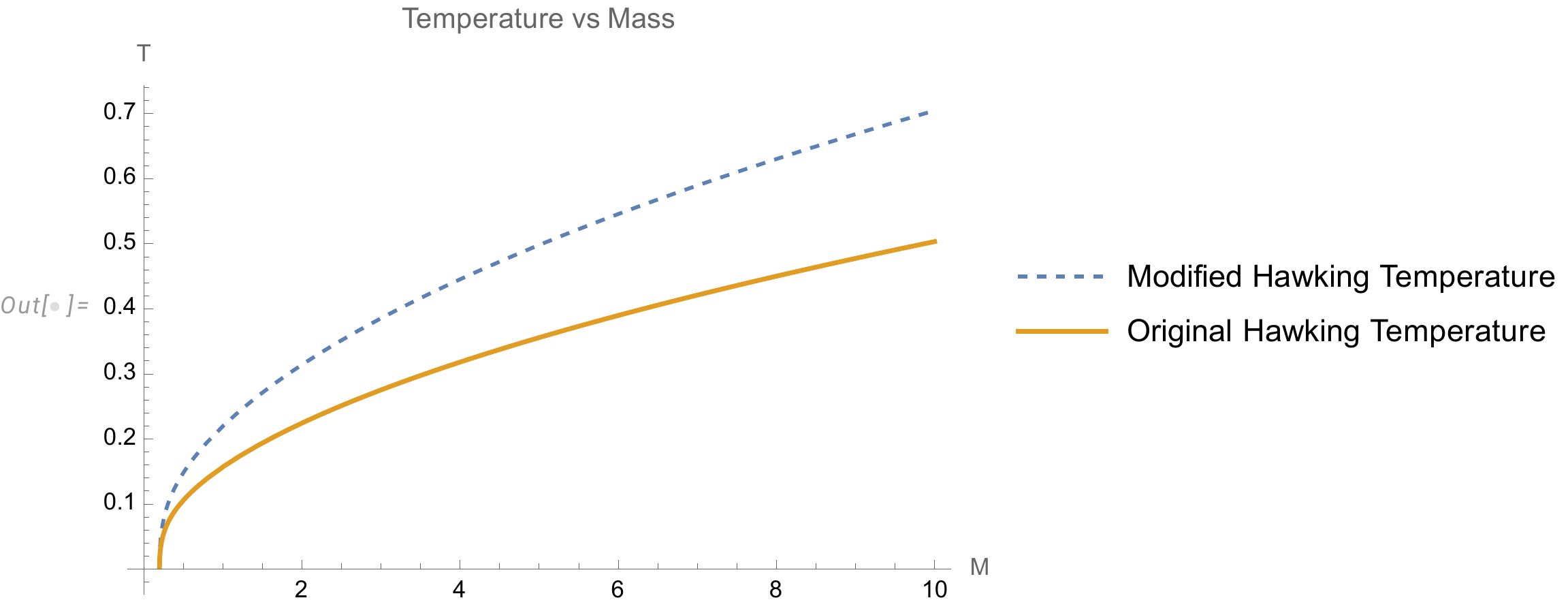}}
\caption{Modified Hawking Temperature vs mass of the black hole by assuming $J=0.2$, $l=1$, $m=0.5$ and $\sigma =0.2$.\label{fig1}}
\end{figure}

\section{Entropy Corrections for rotating BTZ black hole}
In this section, we investigate the correction to the entropy of uncharged rotating BTZ black hole by using the first law of black hole thermodynamics,\cite{26}
\begin{eqnarray}\label{21}
dM = TdS +\Omega dJ,
\end{eqnarray}
where $M$, $T$, $S$, $\Omega$ and $J$ are the mass, the temperature, the entropy, angular velocity and the angular momentum of uncharged rotating BTZ black hole respectively. Using Eq. (\ref{13})  and Eq. (\ref{20}) in Eq. (\ref{21}), we get the modified entropy for rotating BTZ black hole due to modified Hamilton - Jacobi equation as
\begin{eqnarray}\label{22}
S&=& \int \left( \frac{2r_{+}}{\ell^{2}}-\frac{J^{2}}{2r_{+}^{3}}\right) \frac{1}{T_{0}(1+m\sigma \Omega^{2})}dr_{+} ,\cr
&=& S_{BH}-S_{BH}m\sigma \omega^{2}-\frac{1}{S_{BH}}\delta_{1}+S_{BH}(\delta_{2}+\delta_{3})+\frac{1}{S_{BH}}(\delta_{4}-\delta_{5}),
\end{eqnarray}
where $S_{BH}= 4\pi r_{+}$ is the Bekenstein-Hawking entropy of the rotating BTZ black hole and the other coefficients are defined by 
\begin{eqnarray}\label{23}
\delta_{1}&= &\frac{4\pi^{2}J^{2}\ell^{2}}{r_{-}^{2}},\cr
\delta_{2}&=&\frac{\sigma m \omega^{2}J^{2}}{jJ-2r_{+}^{2}\omega},\cr
\delta_{3}&=&\frac{\sigma m \omega^{3/2} J^{3/2}\sqrt{2}}{4\pi\sqrt{j}(jJ-2\omega r_{-}^{2})^{2}}(3jJ+jJ\ell^{2}\omega^{2}-10j^{2}\omega r_{-}^{2}+2\ell^{2}\omega^{3}r_{-}^{2})\frac{1}{2} \log \frac{(\sqrt{2\omega}r_{+}+\sqrt{jJ})}{(\sqrt{2\omega}r_{+}-\sqrt{jJ})},\cr
\delta_{4}&=& \frac{2\pi^{2}r_{+}(J^{2}\ell^{2}-4r_{-}^{4})}{r_{-}^{3}}\log \frac{(r_{+}+r_{-})}{(r_{+}-r_{-})},\cr
\delta_{5}&=& \frac{\sigma m \omega^{4}\pi^{2}r_{-}^{2}r_{+}(J\ell^{2}-4r_{-}^{4})}{jJ-2\omega r_{-}^{2}}4\log \frac{(r_{+}+r_{-})}{(r_{+}+r_{-})}.
\end{eqnarray}
In Eq. (\ref{22}), it is clearly observed that the corrected entropy of rotating BTZ black hole contains inverse term of $S_{BH}$ and logarithmic term of $S_{BH}$. If the coupling constant $\sigma$ sets to zero, the original entropy of the rotating BTZ black hole is recovered as\cite{25}
\begin{eqnarray}
S_o=S_{BH}-\frac{1}{S_{BH}}\delta_1 +\frac{1}{S_{BH}}\delta_4.
\end{eqnarray}
While taking into account the laws of conservation of total energy, total charge and total angular momentum, and on considering the reaction of radiation to the spacetime, the black hole radiation spectrum is found to be non-thermal.$^{6,50}$ This nonthermal spectrum might also be responsible for corrections to the entropy of the black hole.$^{51}$
Counting microstates in string theory and quantum gravity theory not only provided the Bekenstein-Hawking area law but also logarithmic corrections and inverse term in the entropy of a black hole.$^{52-54}$ Also, it is observed that the corrected entropy for rotating BTZ black hole depends on the properties of the emitted fermion particles and also the corrected entropy is related to the angular parameter of the rotating BTZ black hole. This result is consistent with earlier works.$^{38-41}$

\section{Heat capacity of rotating BTZ black hole} 
Another important factor that contributes in the black hole thermodynamics is the heat capacity of the black hole. Heat capacity of a black hole describes the stability of a black hole. The heat capacity of a black hole is\cite{55}
 \begin{eqnarray}\label{24}
C = \frac{\partial M}{ \partial T} = \left( \frac{\partial M}{\partial r_{+}} \right) \left( \frac{\partial r_{+}}{\partial T}\right). 
\end{eqnarray} 
By substituting the black hole mass from $f(r)=0$ i.e, $\frac{\partial M}{\partial r_{+}} = \frac{2r_{+}}{\ell^{2}} - \frac{J^{2}}{2r_{+}^{3}}$ and Eq. (\ref{20}) in Eq. ( \ref{24}), we get
\begin{eqnarray}\label{25}
C = C_{o} (1-m\sigma \Omega^{2} Y),
\end{eqnarray}
where  $C_{o}= \frac{\pi(4r_{+}^{4}-J^{2}\ell^{2})}{r_{+}(r_{+}^{2}+r_{-}^{2})}$ is the original heat capacity of rotating BTZ black hole and 
\begin{eqnarray}\label{26}
Y = \frac{2r_{+}^{2}}{r_{+}^{2}-r_{-}^{2}}-\frac{2jJ}{r_{+}^{2}(\omega -j\Omega)}
\end{eqnarray}
Clearly, if $\sigma = 0$ in Eq. (\ref{25}), the original heat capacity of rotating BTZ black hole is recovered. The modified Hamilton - Jacobi equation reduces the heat capacity of rotating BTZ black hole.  Also, the modified heat capacity also relates to the angular parameter of the black hole and depends on the properties of emitted fermion particles namely mass, energy and angular momentum. Also, positive value of heat capacity shows that the black hole is thermodynamically stable whereas negative value of heat capacity shows that the black hole is thermodynamically unstable \cite{56,57}. Since $\sigma$ is a very small quantity, Eq. (\ref{25}) can never be less than zero. Hence, Eq. (\ref{25}) shows that rotating BTZ black hole is thermodynamically stable black hole. By using Eq. (\ref{25}), a plot between heat capacity and mass of the black hole is presented in Fig. 2 for $m=0.5$ and $\sigma =0.2$. 

\begin{figure}[b]
\centerline{\includegraphics[width=15cm]{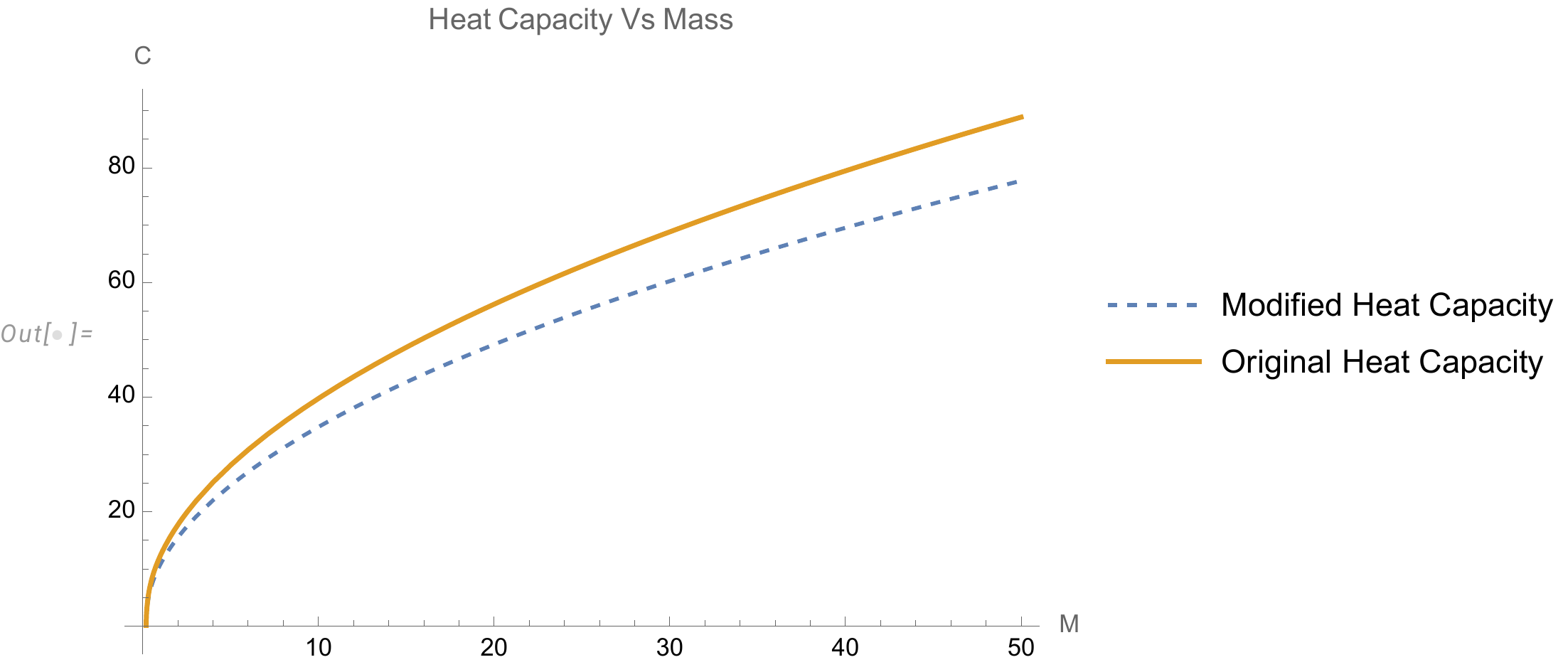}}
\caption{Modified heat capacity vs mass of the black hole by assuming $J=0.2$, $l=1$, $m=0.5$ and $\sigma =0.2$.\label{fig1}}
\end{figure}

\section{Discussions and Conclusions}
In this paper, the thermodynamic properties of rotating BTZ black hole are studied based on MDR and Rarita-Schwinger equation. The modified Hamilton-Jacobi equation is used to investigate the tunneling radiation of fermions across the event horizon of rotating BTZ black hole. It is observed that the modified Hawking temperature of rotating BTZ black hole depends not only on the radial parts of the black hole but also depends on the angular parameter of the black hole. It is also observed that the modified Hawking temperature of rotating BTZ blackhole is more than its original Hawking temperature due to the presence of $\sigma$ term in the deformed Hamilton-Jacobi equation. When $\sigma = 0$, the original Hawking temperature for rotating BTZ black hole is recovered. This result is represented by Fig. 1 by considering $\sigma =0.2$ in Eq. (\ref{20}). 
\\
In the study of black hole thermodynamics, one of the important thermodynamic properties of a black hole is the entropy of the black hole. The corrections to entropy using the modified Hawking temperature is calculated. It is explicitly obtained that the corrections to entropy not only relate to the radial property of the black hole but also on the angular parameter of the black hole. It also depends on the properties of the emitted fermion particles. We can also conclude using Eq. (\ref{20}) and Stefan-Boltzmann law that the fermion emission rate from rotating BTZ black hole computed using Rarita-Schwinger equation is more. The heat capacity of rotating BTZ black hole is investigated. It is observed that both the modified and original heat capacities increase with increase in the mass of the black hole. By using  Eq. (\ref{25}), the variation of heat capacity for a range of mass of the black hole can be seen in Fig. 2 for $\sigma =0.2$. From Eq. (\ref{25}), it is clearly observed that in the absence of deformation parameter $\sigma $, the original heat capacity of rotating BTZ black hole is recovered.

\end{document}